# Application a-Si:H and Its Alloys for Photoreceptor Devices

Darsikin[1], Jasruddin.D.M.[2], H. Saragih[3], D. Rusdiana[4], Sukirno[*], T.Winata[*] and M.Barmawi[*]

[1]Department of Physics, Tadulako University, Palu
[2]Dept. of Physics, Makassar University, Makassar
[3]Department of Physics, Pattimura University, Ambon
[4]Department of Physics, Indonesia Education of University, Bandung
[*]Laboratory for Electronic Material Physics, Dept. of Physics, Bandung Institute of Technology

## Abstract

Photoreceptor devices have been fabricated for positive corona with multi-layer structure of glass/ZnO /i-a-SiC:H /i-a-Si:H /p-a-SiC:H /Al type by using dual chamber plasma enhanced chemical vapor deposition (PECVD) system. 10% Silane ($SiH_4$) and diborane ($B_2H_6$) gases diluted in hydrogen ($H_2$) and 100% methane ($CH_4$) gas were used as gas source. As the top passivation layer, the i-type a-SiC:H with optical band-gap of 2.93 eV was used. This highest optical band-gap was selected to allow more illumination through the photo-conductive layer. The i-type a-Si:H which has optical band-gap of 1.78 eV was used as the photo-conductive layer. The conductivity of p-type a-SiC:H films was also measured. The p-type a-SiC:H film with the lowest dark conductivity was then used as the bottom blocking layer to prevent charge injection from photo-conductive layer to the conductive electrode. The surface voltage characteristic of the device was also investigated. It decays as a function of time, with the maximum surface voltage applied is 120 volts.

## 1. INTRODUCTION

Since the use of hydrogenated amorphous silicon (a-Si:H) material in solar cells the application of this material in opto-electronik devices is carried out intensively. Several devices based on the a-Si:H material for example: photoreceptor, thin film transistor (TFT), thin-film light emitting diode (TFLED) were developed very rapidly.

Photoreceptor which is one of opto-electronic devices was studied and developed since 1970. This device uses in electro-photography process, such as in photocopy machine, facsimile, and laser printer. Several type of organic materials such as: selenium, cadmium sulfida, zincoxide, and amorphous silicon have been studied further for the development of photoreceptor devices. Several structure model have also been developed in order to obtain a good industrial photoreceptor [1],[2].

Compared to other organic materials, a-Si:H has superior characteristic, i. e: high mobility drift, short lifetime, easily controlled optical and electrical properties[3]. This superiority correlates with photoreceptor characteristic, i.e: (1) it can give high charge, $\pm 10^{11}$-$10^{12}$ ion/cm$^2$, with the decay time longer than the process time, (2) it has high surface potential (>200V), and (3) it has good response to light [1].

Most researchers have used a-Si:H thin film and its alloys for fabrication photoreceptor with plasma-CVD

Method[4]. The problem of using this deposition technique is that the dark conductivity of the film is high. This is not suitable when one has to use this film as *bottom blocking layer*, which functions as a layer to prevent charge injection from photoreceptor surface to the electrode.

In this study the results of p-layer a-SiC:H optimization will be described. The fabrication of photoreceptor devices was then carried out based on a-Si:H thin film and its Alloys.

## 2. EXPERIMENTS

The lower dark conductivity of the p-layer a-SiC:H used as the BBL was obtained by optimizing the dopant $B_2H_6$ gas. The deposition parameter was maintained as follow; the $SiH_4$ gas flow rate of 70 sccm, rf frequency of 16.04 MHz, rf power of 30 watt, chamber Pressure of 400 mTorr and distance electrode of 2 cm.

The photoreceptor devices were fabricated following the observation of lower conductivity in p-layer. The devices were fabricated with variation of p-layer a-SiC:H thickness as BBL (1400 Å and 2500 Å) where i-layer a-Si:H thickness as PCL of 6000 Å, and i-layer a-SiC:H as TPL of 500 Å.

Furthermore voltage decay was measured to know the charge storage time in photo-conductivity layer. The measurement method was set by giving bias voltage (using Fluke 5100B) to devices. The Voltage cut off, and the rate of voltage decay was then observed immediately after stopping the bias. The measured voltage was stored in memory of electrometer (Keithley 617) every second. The bias voltage was varied up to a certain value where the devices have no longer responsive. The maximum voltage which can be applied to the devices surface was called surface voltage (Vs).

## 3. RESULTS AND DISCUSSIONS

From The two point probe measurements the results of optimization of the dopant $B_2H_6$ are shown in Figure 1. In this figure the dark conductivity of p-layer a-SiC:H is plotted as a function of ratio $B_2H_6$ gas in a mixture of silane and metane gas. It is clear that conductivity increases as ratio of $B_2H_6$ gas increases. The lowest conductivity of $4.8 \times 10^{-11}$ S.cm$^{-1}$ was obtained when the ratio of $B_2H_6$ gas of 1 %. This film was then selected as BBL in order to minimize charge injection from conductive electrode (Al) to PCL.

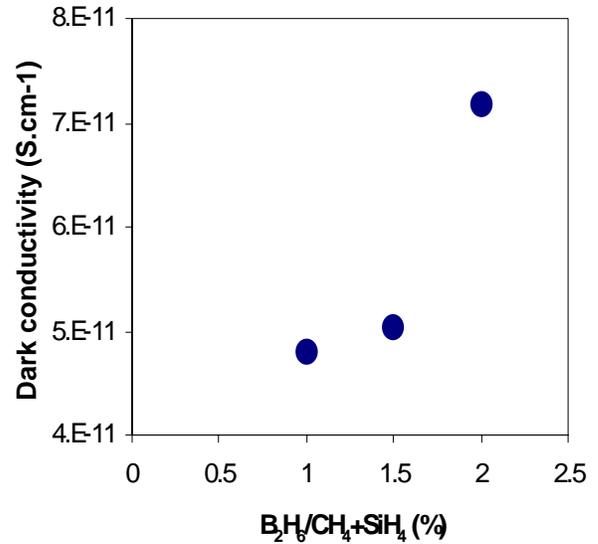

Fig 1. Conductivity of p-layer a-SiC:H shown as the ratio of boron gas in silane and metane ($B_2H_6 /(CH_4+ SiH_4)$) increases

The photoreceptor devices were fabricated by varying the BBL thickness as shown in table 1.

Table 1. The characterization of the grown photoreceptor devices

| Sample | Thickness (Å) | | | Surface Voltage (Vs) |
|---|---|---|---|---|
| | BBL | PCL | TPL | |
| 1 | 1400 | 6000 | 500 | 110 |
| 2 | 2500 | 6000 | 500 | 120 |

Figure 2. shows relation of surface voltage (Vs) versus time of photoreceptor devices in dark and photo induced condition. The Xenon lamp of 24 V, 250 Watt, with intensity of 14,9 mWattcm$^{-2}$ was used during photo induced measurement. It is obvious that the surface voltage tends to decay exponentially. This demonstrates that the photoreceptor devices have a good photo response which implies that the generated carrier has passed through the bulk and BBL

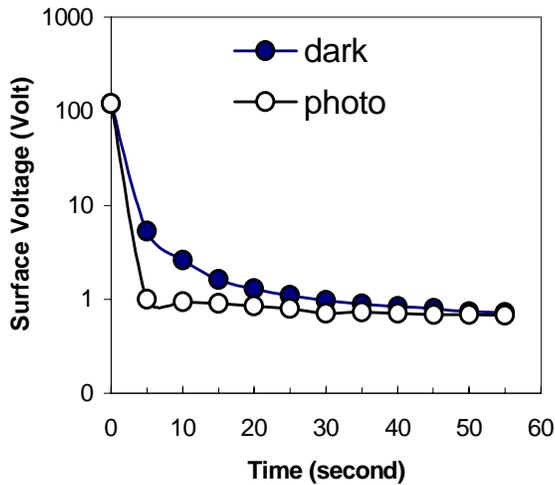

Figure 2. Surface Voltages (Vs) photoreceptor devices as a function of time

## 4. CONCLUSIONS

Surface Voltage of 110 V has been obtained for the photoreceptor devices which have BBL of 1400Å, PCL 6000 Å, and TPL 500 Å. Further higher surface voltage of 120 V was obtained for the photoreceptor devices which have BBL of 2500Å, PCL 6000 Å, and TPL 500 Å. The surface voltage in photo state decays rather fast compared with the dark state. The voltage tends to decay exponentially.

## ACKNOWLEDGMENTS

The authors acknowledge the financial support from CENTER GRANT, RUT-IV, and RUT-V.